\documentclass[aps,twocolumn]{revtex4}
\usepackage{epsfig}
\newcommand{\be}{\begin{equation}}
\newcommand{\ee}{\end{equation}}
\newcommand{\ba}{\begin{array}}
\newcommand{\ea}{\end{array}}

\begin{document}
\title{Effective interactions between star polymers}
\author{Hsiao-Ping Hsu and Peter Grassberger}
\affiliation{John-von-Neumann Institute for Computing, Forschungszentrum
J\"ulich, D-52425 J\"ulich, Germany}


\date{\today}

\begin{abstract}
We study numerically the effective pair potential between two star polymers
with equal arm lengths and equal number $f$ of arms. The simulations were 
done for the soft core Domb-Joyce model on the simple cubic lattice, to 
minimize corrections to scaling and to allow for an unlimited number of arms.
For the sampling, we used the pruned-enriched Rosenbluth method (PERM).
We find that the potential is much less soft than claimed in previous
papers, in particular for $f\gg 1$. While we verify the logarithmic
divergence of $V(r)$, with $r$ being the distance between the two cores,
predicted by Witten and Pincus, we find that the Mayer function 
for $f>20$ is hardly distinguishable from that for a Gaussian potential.
\end{abstract}

\maketitle 

Interactions between polymers in diluted solutions are of interest
for several reasons, not the least because they influence both the 
equilibrium and the rheological properties of complex fluids. In early
work by Flory {\it et al.}~\cite{flory} it was suggested that polymer 
coils can be approximated by hard spheres, but this was shown to be 
wrong in \cite{grosberg}. Since then it is well understood that both 
linear and branched polymers are {\it soft} in the sense that they can 
penetrate each other, and that the effective potential is a rather smooth 
function of their distance. As shown in \cite{lowen0,lowen}, this can
have dramatic effects on the phase diagram for semi-dilute solutions 
of star polymers, and can -- with the effective potentials assumed 
by these authors -- lead to a multitude of novel phases.

When discussing effective potentials between polymers -- be they
linear or star-shaped -- one has to distinguish between $U(r)$ where 
$r$ is the distance between the two centers of mass, and $V(r)$ 
where $r$ is the distance between the two central monomers. In both 
cases, the potential is defined by 
\be
   \exp(-\beta U(r)),\; \exp(-\beta V(r)) = Z^{(2)}(r)/[Z^{(1)}]^2   \label{Z2f}
\ee
where $Z^{(1)}$ is the partition function of a single polymer, while 
$Z^{(2)}(r)$ is the partition function of two polymers with fixed distance 
$r$.  Finally, $\beta = 1/k_BT$ is used to give $V(r)$ the usual dimension
of a potential, although any temperature is of course dummy for an
a-thermal system as in the present case. For ease of writing, we shall 
set $\beta=1$ in the following. Finally, all data shown in the following 
refer to lattice simulations with ${\bf r} = (r,0,0)$, 
but we checked in a few cases that
distances not parallel to one of the coordinate axes gave basically the 
same results.

In the following we shall only discuss the case where the number $f$ of 
arms is the same for both polymers (and might include the case $f=2$ 
describing linear chains), and each arm has the same length $N$. Even if 
not spelled out explicitly, the main point of \cite{lowen0,lowen,Likos}
is that, for large $f$, the potential $V(r)$ is more relevant than $U(r)$ 
for equations of state of semi-dilute or dense solutions, and that $V(r)$
is very different from $U(r)$: While the latter is essentially Gaussian,
the former has a more complex structure with a Yukawa tail at large $r$.
We will show in the following that at least the second claim is not correct, 
and that $V(r)$ can also be approximated by a Gaussian for most practical
purposes.

The center-of-mass potential $U(r)$ is well known to be approximately
Gaussian for linear \cite{grosberg,kruger,dautenhahn,bolhuis} polymers.
For star polymers there are much fewer computations \cite{zifferer}, 
so we present in Fig.~1 our own results which clearly indicate that
$U(r)$ is roughly Gaussian, too. Notice that the deviation from a Gaussian
at small $r$ (i.e. the upward bending in Fig.~1) is practically irrelevant 
for $>8$ arms per star, since it occurs only when $e^{-U(r)} \leq 10^{-3}$.

\begin{figure}
   \psfig{file=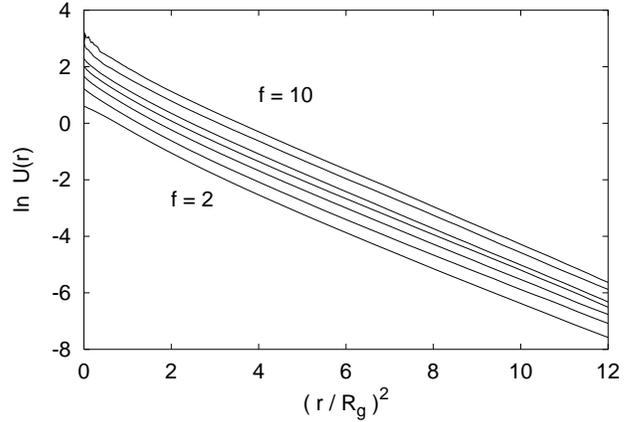,width=6.0cm, angle=270}
   \caption{Logarithms of the effective potential $U(r)$ with $r$ 
      being the center-of-mass distance,
      for star polymers with $f=2, 3, 4, 5, 6, 8,$ and $10$ arms (bottom to
      top), plotted against $(r/R_g)^2$ where $R_g$ is the gyration 
      radius of a single star. Arm lengths are $N=400$ monomers.
      Linear curves would correspond to Gaussian $U(r)$.}
\label{U_r}
\vspace{-2mm}
\end{figure}

The data in Fig.~1 as all data in this paper were obtained for the 
soft repulsion Domb-Joyce model \cite{domb} at the `magic' value 
$v^*=0.6$ 
of the repulsion parameter, on the simple cubic lattice. This model 
was chosen because it leads to minimal corrections to scaling and it 
allows an arbitrary number $f$ of arms to be attached to a single central 
site \cite{star,footnote1} 
The simulations were made with the PERM algorithm 
\cite{perm}, adapted for star polymers as described in \cite{star}. The 
partition sum $Z^{(2)}(r)$ was estimated as usual (e.g. \cite{dautenhahn}) 
by simulating two {\it independent} stars simultaneously, and computing their 
overlaps at different distances.

From general scaling arguments we expect $U(r)$ and $V(r)$ to depend on 
the arm length $N$, for $N\gg 1$, only via the scaling variable $x = r/R_g$, 
where $R_g$ is the gyration radius of the star (for large $N$, $R_g$ 
scales as $R_g \approx \sqrt{A_f} N^\nu$, with values of $A_f$ given 
in Table~1). We checked this by making plots similar to Fig.~1 also for 
other values of $N$ (not shown here) and by estimating $U(0)$ for different
$N$. As argued in \cite{kruger,bolhuis} for linear polymers ($f=2$), the 
convergence for $N\to\infty$ is from above, $U_{N,f}(0) \approx 
U_{\infty,f}(0) + a/N^{0.7}$, for small $f$ ($f\leq 6$); for larger $f$ the 
data were ambiguous.  
For $f=2$ and 4, which are the only cases where precise comparisons to 
previous work are possible, the data shown in Fig.~1 are in perfect agreement 
with \cite{bolhuis,zifferer}. Values of $U(0)$, extrapolated to $N\to\infty$, 
are also given in Table~1. They seem to scale as $U(0)\approx 0.6 f^{1.58}$.

\begin{table}
\begin{center}
\caption{Main results. The numbers in brackets are single standard deviations 
  in the last digit. $A_f$ is defined by $R_g^2 \approx A_f N^{2\nu}$, $b_f$ is 
  obtained through Eq.~(\ref{bf}) from the data of \cite{star}, 
  $U(0)$ is the effective potential when the two centers of mass
  coincide, and $a_f$ is defined in Eq.~(\ref{wp}), $c_f$ and $d_f$ are defined
  in Eq.~(\ref{gauss}), and $\tau_f$ is defined in Eq.~(\ref{Vnew}). We do not
  quote errors for the latter four, since they are strongly correlated and 
  individual error estimates would not make sense.}
  \label{table1}
\begin{tabular}{r|l|c|c|l|c|l|c}
  $f$& $\;\;A_f$ & $b_f$    & $U(0)$   & $\;a_f$ &   $c_f$  & $\;d_f$ & $\tau_f$ \\ \hline
   2 & 0.2902(2) & 0.815(2) & 1.791(2) & 1.869   & 0.372    & 0.405   &  4.5     \\
   3 & 0.3587(2) & 1.540(3) & 3.357(6) & 1.759   & 0.74     & 0.473   &  2.2     \\
   4 & 0.4017(2) & 2.415(5) & 5.11(2)  & 1.720   & 1.17     & 0.506   &  1.35    \\
   5 & 0.4337(3) & 3.42(1)  & 7.27(4)  & 1.707   & 1.76     & 0.527   &  1.00    \\
   6 & 0.4596(4) & 4.52(2)  & 9.60(11) & 1.682   & 2.90     & 0.548   &  0.98    \\
   8 & 0.5008(5) & 7.05(2)  &15.9(4)   & 1.690   & 4.62     & 0.582   &  0.62    \\
  10 & 0.5343(6) & 9.90(3)  &23.2(11)  & 1.691   & 7.0      & 0.600   &  0.50    \\
  12 & 0.5629(8) &13.15(6)  &34.(4)    & 1.70    & 10.6     & 0.610   &  0.53    \\
  14 & 0.588(1)  &16.71(8)  &    -     & 1.71    & 14.1     & 0.62    &$\approx 0.6$\\
  16 & 0.612(2)  &20.54(10) &    -     & 1.67    & 19.0     & 0.65    &$\approx 0.6$\\
  18 & 0.632(2)  &24.73(14) &    -     & 1.69    & 22.5     & 0.65    &$\approx 0.5$\\
  20 & 0.652(2)  &29.3(2)   &    -     & 1.73    & 26.      & 0.64    &$\approx 0.5$\\
  24 & 0.689(3)  &39.7(3)   &    -     & 1.76    & 39.      & 0.65    &$\approx 0.7$\\
  30 & 0.735(3)  &57.3(6)   &    -     & 1.75    &  54.     & 0.67    &$\approx 0.7$\\
  35 & 0.764(4)  &76.3(11)  &    -     & 1.78    &  76.     & 0.68    &$\approx 0.5$\\
  40 & 0.790(4)  &94.6(20)  &    -     &    -    &    -     &   -     &   -     \\
  50 & 0.846(5)  &    -     &    -     &    -    &    -     &   -     &   -     \\
  60 & 0.870(7)  &    -     &    -     &    -    &    -     &   -     &   -     \\

\end{tabular}
\end{center}
\vspace{-2mm}
\end{table}

\begin{figure}
   \psfig{file=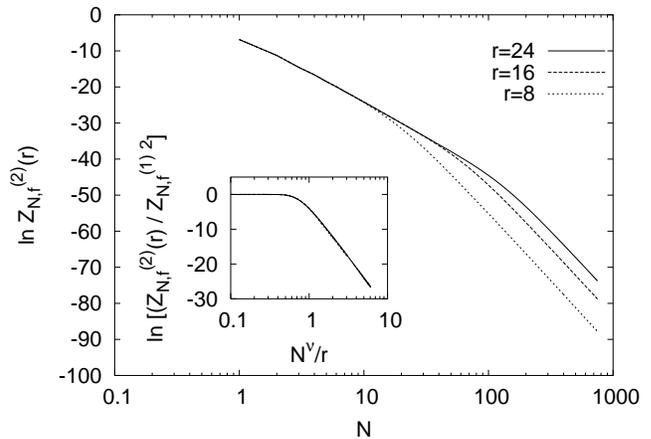, width=6.0cm, angle=270}
   \caption{Logarithms of partition functions $Z_{N,f}^{(2)}(r)$
     against $N$ for $f=12$. The data collapse expected from the cross-over
     ansatz Eq.~(\ref{ZN2fa}) is shown in the insert.}
                \label{lz12}
\vspace{-2mm}
\end{figure}

Much more attention had been given in the literature previously to the 
potential $V(r)$ with $r$ being the central monomer distance, and we shall
also concentrate on $V(r)$ in the following. The first result on it
was obtained by Witten and Pincus~\cite{witten}. They pointed out that 
the scaling \cite{dupla}
\be
   Z_{N,f}^{(1)} \sim \mu^{-fN} N^{\gamma_f-1} \; , \label{ZNF}
\ee
of the partition sum of a star with $f$ arms and arm length $N$, together
with the assumption that $Z_{N,f}^{(2)}(r) / [ Z_{N,f}^{(1)}]^2$ is for 
any fixed $f$ a function of $x\equiv r/R_g$ only,
\be
   Z_{N,f}^{(2)}(r) / [Z_{N,f}^{(1)}]^2 = \psi_f(r/R_g)\;,   \label{ZN2fa}
\ee
implies that
\be
   V(r) \approx V_{\rm WP}(r) \equiv  b_f \ln (a_f R_g/r)   \label{wp}
\ee
for $1 \ll r \ll R_g$ with 
\be
   b_f=(2\gamma_f-\gamma_{2f}-1)/\nu\;.              \label{bf}
\ee
Precise estimates of $\gamma_f$
can be found in~\cite{star}. They show that the scaling $b_f\sim f^{3/2}$
obtained in \cite{witten} by assuming the phenomenological Daoud-Cotton 
model \cite{daoud} is not exact, a power law fit gives instead $b_f\approx 
0.27f^{1.58}$. Both $a_f$ and $b_f$ should be universal
and should not depend on the specific microscopic realization.

This is illustrated in 
Fig.~2 where we show $\ln Z_{N,f}^{(2)}(r)$ as a function of $N$, for
$f=12$ and for three different values of $r$. In contrast to the data shown 
in Fig.~1, these data were obtained by growing the two stars at distance 
$r$ and {\it with the mutual interactions taken into account} during the 
growth \cite{footnote2}.
This allows to measure $Z_{N,f}^{(2)}(r)$ down to
very small distances and large $N$, where it is so small that the ratio
$Z_{N,f}^{(2)}(r) / [Z_{N,f}^{(1)}]^2$ measured from independently grown 
stars would be indistinguishable from zero. On the other hand, at large 
distances this second method would give very bad estimates of $V(r)$, 
since it is obtained by subtracting the (nearly equal) free energies 
obtained in two independent runs. Therefore, in the following, all 
plots will show data obtained either by the first or by the 
second method, or will contain combinations of both types of data.

\begin{figure}
   \psfig{file=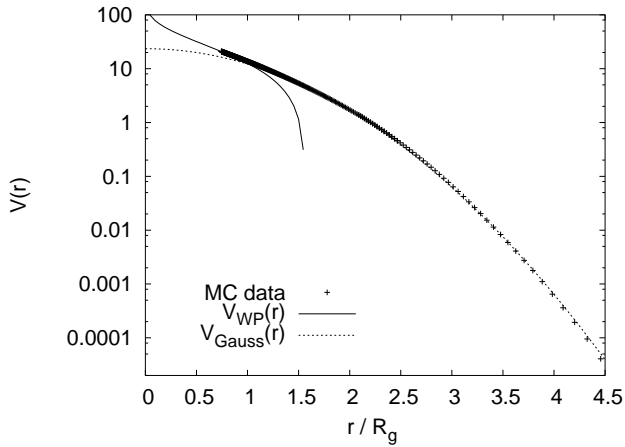, width=6.0cm, angle=270}
   \caption{$V(r)$ for $f=18$, plotted on a logarithmic scale against $r/R_g$. 
   The short continuous curve on the left corresponds to Eq.~(\ref{wp}), 
the dashed curve is a Gaussian.}
\label{V18}
\vspace{-2mm}
\end{figure}

Eq.~(\ref{wp}) cannot hold for large distances, and it is there where
previous results were most uncertain. An analytic ansatz which is supposed 
to cover all values of $r$ was made by Likos et al.~\cite{Likos}. Using 
a `corona' radius $\sigma$ \cite{witten} which is roughly comparable 
in size to the gyration radius, they assumed that 
\begin{eqnarray}
  V(r)=  & & \nonumber \\
     \frac{5 f^{3/2}}{18} & 
        \left\{\begin{array}{ll}
           -\ln(r/\sigma)+\frac{1}{1+\sqrt{f}/2} & {\rm for}\; r \leq \sigma \\
           \frac{\sigma/r}{1+\sqrt{f}/2}\exp\left(-\sqrt{f}(r-\sigma)/2\sigma
         \right) & {\rm for}\; r > \sigma
        \end{array} \right . \label{VLoewen}
\end{eqnarray}
This was supported by molecular dynamics simulations and was also shown to 
be compatible with experimental results. It was used in extensive simulations 
of semi-dilute and concentrated solutions, and gave rise to a number 
of very interesting predictions~\cite{lowen0,lowen}. But for 
linear polymers it disagrees with the analytic results of \cite{kruger} 
and seems hard 
to be reconciled with the simulations of~\cite{kruger,bolhuis,zifferer}.
In particular, it was shown in~\cite{Freire} that Eq.~(\ref{VLoewen}) is
in gross violation with simulations of off-lattice stars with up to 18
arms. But these arms were very short, whence one might doubt the relevance
of the results of~\cite{Freire}. 

Anyhow, in a later paper Jusufi {\it et al.}~\cite{jusufi2} proposed to 
use Eq.~(\ref{VLoewen}) only for $f>10$, and to replace it for $f<10$ by
an ansatz with Gaussian decay for $r>\sigma$,
\begin{eqnarray}
  V(r)= & & \nonumber \\
     \frac{5}{18} &  f^{3/2}
        \left\{\begin{array}{ll}
           -\ln(r/\sigma)+{1\over 2\tau^2\sigma^2}& {\rm for}\; r \leq \sigma \\
           {1\over 2\tau^2\sigma^2} \exp(-\tau^2(r^2-\sigma^2)) & {\rm for}\; r > \sigma
        \end{array} \right . \label{VLoewen2}
\end{eqnarray}
Notice that this does not alleviate the serious conflict with~\cite{Freire}.
Also, we would expect that the center of mass gets closer to the central monomer
as $f$ increases. Thus, if $U(r)$ is roughly Gaussian for large $r$, we should 
expect that also $V(r)$ is Gaussian there for $f\gg 1$.

Let us for the moment concentrate on $f=18$ arms, the case studied in \cite{Likos}.
In order to get a first overall impression of $V(r)$, we show in Fig.~\ref{V18} 
its logarithm, obtained for fixed $r=20$ and for all $N \leq 400$, against $r/R_g$.
The short continuous curve at small r is the Witten-Pincus prediction, modified
by taking the measured values of $\gamma_f$ and $\gamma_{2f}$. It is relevant only 
for $r\ll R_g$. For $r\gg R_g$ the MC data can be approximated by a parabola, i.e. 
$V(r)$ is roughly Gaussian,
\be
   V(r) \approx V_{\rm Gauss}(r) \equiv c_fe^{-d_fr^2/R_g^2}\;.   \label{gauss}
\ee
We conjecture that $c_f$ and $d_f$ are universal.
A Yukawa tail as in Eq.~(\ref{VLoewen}) would essentially correspond to a straight 
line in Fig.~\ref{V18} and is definitely ruled out~\footnote{Actually,
$V(r)$ decays for $r\rightarrow \infty$ {\it faster} than Gaussian,
as $V(r)\sim e^{-r^\delta \cdot const}$ with $\delta=(1-\nu)^{-1}>2$
~\cite{deGenn}. This follows from the fact that arms are very far
from each other for $r \gg R_g$, and thus the potential is proportional to
the product of the densities
in a single unbranched chain~\cite{deGenn}.
But we expect this to hold only for very large $r$, far beyond the distances
we could study in this paper.}. 

\begin{figure}
\psfig{file=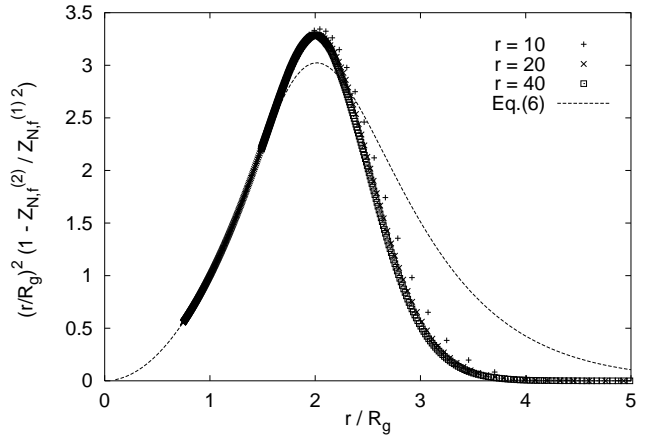, width=6.0cm, angle=270}
   \caption{Rescaled radial Mayer functions against $r/R_g$ for $f=18$.
     The dashed curve is the prediction from Eq.~(\ref{VLoewen}) with 
     $\sigma=1.3R_g$~\cite{Jusufi}.}
\label{lm2f18b}
\vspace{-2mm}
\end{figure}

Since $r=20$ is not very large, one might be worried about finite size corrections. 
When plotted as in Fig.~\ref{V18}, finite size corrections would be visible only
in the r.h.s. tail where $V$ is so small that they are irrelevant. Thus we plotted 
in Fig.~\ref{lm2f18b} the rescaled radial Mayer function,
\be
   (r/R_g)^2f_M(r)=(r/R_g)^2(1-\exp(-V(r))
\ee
which is the most 
interesting quantity, for three values of $r$. This plot agrees very well with the 
simulations of~\cite{Freire}, although those authors used a continuum model with 
much shorter arms. On the other hand, our data disagree strongly with 
Eq.~(\ref{VLoewen}) which is indicated by the dashed curve.

\begin{figure}
   \psfig{file=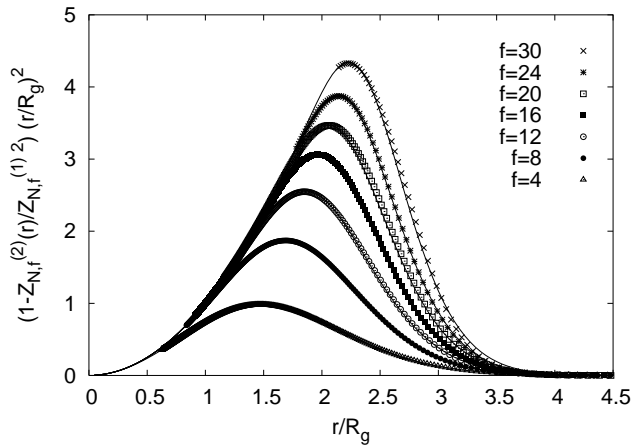,width=6.0cm, angle=270}
   \caption{Rescaled radial Mayer functions against $r/R_g$ for several values
    of $f$. Curves are obtained from Eq.~(\ref{Vnew}), with fitted parameters
   $a_f, c_f, d_f$ and $\tau_f$ given in Table~1.}
\label{lmall}
\vspace{-2mm}
\end{figure}

Linking the small- and large-$r$ behaviours seen in Fig.~\ref{V18} into a piecewise 
analytic form as in Eqs.~(\ref{VLoewen}) or (\ref{VLoewen2}) would obviously 
give a discontinuous slope and a bad fit. Rather we found that 
the following ansatz  
describes all our data quantitatively, for all $2\leq f<35$ and for all values of $r$:
\be
   V(r)= {1\over\tau_f} \ln\left[e^{\tau_fV_{\rm WP}(r)-d_fr^2/R_g^2}+e^{\tau_fV_{\rm Gauss}(r)}\right]\;. 
                  \label{Vnew} 
\ee
with $V_{\rm WP}(r)$ and $V_{\rm Gauss}(r)$ defined in Eqs.~(\ref{wp}) and (\ref{gauss}),
and with $\tau_f$ being an additional parameter for every $f$. It is easy to see that
$V(r)>0$ for all $r$ and that $V(r)= V_{\rm Gauss}(r)\;[1+O(r^{-b_f})]$ for 
$r\to\infty$, while $V(r)= V_{\rm WP}(r)\;[1 + O(r^2)]$ for $r\to 0$. Like the previous
parameters, also $\tau_f$ should be universal.
Values for $a_f, c_f, d_f$, and $\tau_f$ obtained by fitting our MC simulations are 
given in Table~1. One sees that $\tau_f$ is between 1/2 and 1, except for the smallest
values of $f$. The strength of $V_{\rm Gauss}(r)$ increases roughly as $c_f \approx 0.1f^{1.88}$. 
Its range increases faster than $R_g$ and the peak of the radial Mayer function
increases even faster, roughly as $R_g\ln f$. For several values of $f$, radial Mayer 
functions are shown in Fig.~\ref{lmall} together with the fits obtained with 
Eq.~(\ref{Vnew}). For $f\gg 1$, their peaks are at $r/R_g > a_f$,
i.e. at distances where $V_{\rm WP}(r)$ would be negative.

For $f>20$ our ansatz for $V(r)$ can be simplified. For such stars the potential 
is so big for small $r$ that the Witten-Pincus term can be neglected for dilute
solutions: Whenever it would be relevant in comparison to the Gaussian term, the 
pair distribution function $\exp(-V(r))$ is already zero for all practical 
purposes~\cite{footnote3}.
The Witten-Pincus part becomes important only for 
very dense systems. But there the description in terms of effective two-body
forces is questionable. For the same reason, also the parameter $\tau_f$ is less 
precisely determined than $a_f$, $c_f$ and $d_f$.

In summary, we have obtained very precise Monte Carlo estimates of the effective
potentials between two star polymers with equal number of arms and equal arm 
lengths. Using a soft core polymer model (the Domb-Joyce model) we have 
reduced corrections to scaling to a minimum, and we have been able to 
simulate many arms without having to use a large central particle. We thus 
believe that our results present essentially the scaling limit of long 
arms. Our most important finding is that effective potentials are much 
harder than previously believed. This refers to the case where the central
monomers are used to define the distance. For the alternative case of the 
center-of-mass distance, it had already been assumed by previous authors that 
the potential is relatively hard at large $r$ and approximately
Gaussian. We found that basically the same is true also for the central
mass definition. Which of these two alternatives is a better starting point
for effective potentials in systems with finite concentration is another
question, but our results suggest that it does not make much difference.

We thank Walter Nadler for discussions and for critically reading the manuscript,
and Christian von Ferber for interesting correspondence.


\begin{thebibliography}{30}
\bibitem{footnote1} In \cite{star} we had considered two versions, one where
   the center of the star is occupied by a single monomer and one with $f$-fold
   occupation. In the present paper we calculated centers of mass and gyration
   radii as if the centers were singly occupied (else, the center would have too   large a weight). But we computed Boltzmann weights as if the centers were
   $f$-fold occupied. The latter has little effect on the outcome, except that
   it makes the simulations more efficient.
\bibitem{footnote2} Notice that this strategy cannot be used for calculating $U(r)$,
   since the center of mass distance is not fixed even if the central monomers have
   a fixed distance.
\bibitem{footnote3} This might not be true if there are chemically reactive
   radicals at the star centers. In that case even a very small probability for
   the star centers to meet might be important, and the difference between
   Eqs.~(\ref{Vnew}) and the purely Gaussian approximation might be relevant.
\end{thebibliography}
\end{document}